\documentclass[sigconf]{acmart}
\usepackage[toc,page]{appendix}
\usepackage{enumitem}
\renewcommand\footnotetextcopyrightpermission[1]{} 
\usepackage{booktabs} 

\setcopyright{rightsretained}

\acmConference[ACM FAT\*]
\acmYear{2019}
{Atlanta, USA}

\author{Gizem Gezici}
\affiliation{%
  \institution{Sabanci University, Istanbul, Turkey}
  \institution{Huawei Turkey R\&D Center, Istanbul, Turkey}}
\title{Biased or Not?: The Story of Two Search Engines}

\begin{document}

\begin{abstract}
Search engines can be considered as a gate to the world of WEB, and they also decide what we see for a given search query. Since many people are exposed to information through search engines, it is fair to expect that search engines should be neutral; i.e. the returned results must cover all the elements or aspects of the search topic, and they should be impartial where the results are returned based on relevance. However, the search engine results are based on many features and sophisticated algorithms where search neutrality is not necessarily the focal point. In this work we performed an empirical study on two popular search engines and analysed the search engine result pages for controversial topics such as abortion, medical marijuana, and gay marriage. Our analysis is based on the sentiment in  search results to identify their viewpoint as conservative or liberal. We also propose three sentiment-based metrics to show the existence of bias as well as to compare viewpoints  of the two search engines. Extensive experiments performed on controversial topics show that both search engines are biased, moreover they have the same kind of bias towards a given controversial topic.
\end{abstract}


\maketitle

\section{Introduction}

Search engines are a major source of information for many people especially in the new millennium. While early search engines such as Alta Vista had ranking mechanisms which were fairly simple and transparent, later developments such as Google's page rank algorithm and further ranking  methods based on click streams completely changed that situation. Current search engines perform retrieval and ranking based on many features introducing a complexity way beyond the simple model of relevance used in early search engines.   

The underlying ranking mechanisms of search engines indirectly determine the information people are  exposed regarding various topics including controversial ones. On the other hand, due to the fact that search engines are driven by algorithms, general public may expect that they are neutral in the sense that the returned results are expected to cover all the elements or aspects of the search topic, and they should be impartial where the results are returned based on relevance. However, search neutrality is not necessarily the focal point of modern search engines\footnote{https://plato.stanford.edu/entries/ethics-search/\#SearEngiBiasProbOpac}.
Given a search engine through which users are exposed to the news coming from different sources where the Search Engine Result Pages (SERPs) are gathered and ranked by the search engine, there may be a bias towards a view point in the result set. Assume that the user searches for \textit{2016 Presidential Election} and the top-k ranked results are displayed to the user. The resulting set may be biased towards a specific political view, such as \textit{conservative} or \textit{liberal}.  Defining bias in a set of documents is challenging by itself, and the issue becomes even more complicated if ranking is taken into account since we are not only interested in the distribution of positive and negative opinion in the result set but also the ranking in the distribution. Current research in ranking employs learning to rank methods to return the most relevant results (i.e. web documents) for a requested query to the user. So the bias may come from the design of the ranking algorithm, i.e., training set of the learning to rank algorithms such as click-through logs which intrinsically contain \textit{user bias}, or machine learning features proposed by algorithm designers which inherit \textit{human-like tendencies} that lead to human bias to train the system. When personalisation is involved together with learning to rank algorithms, then the bias can be justified by the fact that the resulting set of documents are determined based on the user preferences. For instance, if the user is inclined towards liberal news which is implied by her previous clicks on the returned results then it is expected that the system learns that the user prefers positive news (or documents) regarding the liberals and presents the user with top-k liberally biased documents. Otherwise, the presented results have what we call ''unjustified bias'' towards a view point. People seriously rely on the results of search engines and they are influenced by returning documents for a given topic. 
Therefore, unjustified bias is becoming more dangerous, if users query strongly polarized (i.e. controversial) topics. Hence, unjustified bias in retrieved results especially for controversial queries is our main focus.

We analyse the SERPs of two major search engines, Google and Bing, in order to detect any significant difference in the viewpoints reflected in SERPs. We utilize opinion mining to understand if a search result page has a positive or a negative bias towards a specific view point. We collected and analysed the SERPs by at  two popular search engines to ensure that the bias is not specific to one particular search engine. In this work, we try to answer the following research questions:
 \begin{itemize}
\item\textbf{RQ1: } Do search engines return \textit{non-biased} result sets in response to queries related to controversial issues? \\                    
\item\textbf{RQ2: } On conservative-liberal scale, are the viewpoints of search engines \textit{significantly different} from each other towards controversial topics?
\\                    
\item\textbf{RQ3: } If the retrieved results of search engines include bias, can we determine \textit{the source of bias} as input bias or algoritmic bias?
 \end{itemize}

Given the above two research questions, in the scope of RQ1, detecting unjustified bias is a challenging problem especially when ranking is considered. This is because, the bias cannot be measured only by the distribution of the sentiment in the result set. The bias measure should incorporate the rank of the returned results, as well. For the second question, we compare the two search engines by taking into account sentiment as well as ranking information and see if they show similar/different viewpoints on a given controversial topic.

More specifically, we propose a sentiment-based comparison procedure to understand if the retrieved documents of search engines contain bias, and also to make a pairwise comparison of their perspectives on controversial issues. Experimental results demonstrate that both search engines return \textit{biased} results for our sentiment-based metrics. However, the bias may originate
from the corpus itself which is known as "input bias". 
The corpus 
Moreover, we show that Bing and Google have similar perspectives on the corresponding controversial topics.


The paper is organised as follows. We first provide related work on search bias. Second, we present our framework by explaining our sentiment-based comparison metrics in detail. Then, we provide details regarding our SERP dataset crawling and evaluation framework. We report our experimental results and discuss our findings, and finally, we conclude our paper.
 

\section{Related Work}

The term "search neutrality" is explained in a NYTimes article as "The principle that search engines should have no editorial policies other than that their results be comprehensive, impartial and based solely on relevance \footnote{https://www.nytimes.com/2009/12/28/opinion/28raff.html}

Issues regarding search engine bias have been summarized as ~\cite{tavani2012search}:
\begin{itemize}
\item search-engine technology is not neutral, but instead has embedded features in its design that favor some values over others; 
\item major search engines systematically favor some sites (and some kind of sites) over others in the lists of results they return in response to user search queries; and 
\item search algorithms do not use objective criteria in generating their lists of results for search queries.
\end{itemize}

A study done by Robert Epstein and Ronald E. Robertson demonstrated the effect of search engines in elections ~\cite{epstein2015search} The search engine manipulation effect (SEME) and its possible impact on the outcomes of elections. Here is an excerpt from the abstract "We present evidence from five experiments in two countries suggesting the power and robustness of the search engine manipulation effect (SEME). Specifically, we show that (i) biased search rankings can shift the voting preferences of undecided voters by 20 percent or more, (ii) the shift can be much higher in some demographic groups, and (iii) such rankings can be masked so that people show no awareness of the manipulation." Their work is based on user studied from different countries, and they do not have techniques to automatically detect specific bias and ranked results. In a more recent work by the same researchers, they presented that the existence of bias in election-related search results can have a strong and undetectable influence on the preferences of undecided voters ~\cite{epstein2017suppressing}. This work also does not include a method of automatically identifying search bias, rather an empirical study.

Here is a blog that reports a study on "Google bias in search results; 40 percent lean left or liberal" \footnote{(http://www.canirank.com/blog/analysis-of-political-bias-in-internet-search-engine-results/)}. They relied on human labelers from different political spectrum. Another popular article in the Guardian states that "Google's search algorithm appears to be systematically promoting information that is either false or slanted with an extreme rightwing bias on subjects as varied as climate change and homosexuality." \footnote{https://www.theguardian.com/technology/2016/dec/16/google-autocomplete-rightwing-bias-algorithm-political-propaganda}

Our main focus in this work is identifying and measuring the degree of bias in Search Engine Result Pages (SERPs) with an emphasis on news sources such as NYTimes, and BBC news. 

For example based on the analysis performed on a popular search engine, Yom-Tov et. al. in~\cite{yom2014promoting} show that people are in fact more likely to read opinions matching with their own and people are more likely to read news from opposing sites given that the language model of a particular news item is close to the language model of their own political view. Yom-Tov et. al. also show that people who were presented with more diverse results continued reading more diverse results and overall became more interested in news.

Related research areas to Bias in Search Engines are: \textit{Topic Discovery from text}, \textit{Sentiment Analysis}, \textit{Echo Chambers} and \textit{Filter Bubbles}. Topic discovery and sentiment analysis are essential to automatically identify if the results of a query provide information about all the aspects and opinions on those aspects, thereby avoiding any bias on search results for a specific topic. 

Fang et. al. proposed a novel opinion mining research problem, which they call Contrastive Opinion Modeling (COM)~\cite{fang2012mining}. For a given query topic over a database of text collections representing multiple perspectives, Fang et.al.~\cite{fang2012mining}, define the task of COM as presenting the opinions of the individual perspectives on the query topic. Authors also quantify the difference of perspectives by the Jensen-Shannon divergence among the individual topic-opinion distributions. In our framework, we observed that topic discovery may work for comparing search engines, however it ,s not sufficient for the fairness analysis of search engines. For this reason, we decided to use sentiment instead of topic information to compare the perspectives of the given search engines and measure the fairness in the retrieved documents.



Aktolga and Allan consider the sentiment towards topics and propose different query diversification methods based on the topic sentiment~\cite{aktolga2013sentiment}. They also measure the stability of the sentiment classification with various accuracy levels. A parallel work by Naveed et. al. is related to "Feature Sentiment Diversification of User Generated Reviews: The FREuD Approach"~\cite{naveed2013feature}.

More similar work to ours was fulfilled by Kulshrestha et. al. ~\cite{kulshrestha2017quantifying} in which they measured different types of bias in search engines, i.e., input or source bias, ranking system bias, and output bias which is the cumulative of these two. They proposed different metrics to compute and differentiate the distinct types of bias in political-related tweets. Our work is different from their work mainly in four ways. Firstly, we used news documents which contain structural textual content not tweets for a detailed analysis. Secondly, we did not propose new metrics, rather we utilized commonly used traditional IR metrics (NDCG and average precision) by only modifying relevance grades with sentiment polarity values to detect bias. Thirdly, we evaluated the retrieved documents of two commercial search engines instead of Twitter and compared their perspectives, as well in addition to solely detecting bias. And lastly, we did not measure only political bias, but investigated the degree of bias (i.e., by using a fairness metric) in search engines for many diverse controversial topics-related queries.

In addition to the search engine ranking system bias, the bias may also be resulted from query, if the query itself has some polarity or supports a specific viewpoint on a given topic. Things tend to be phrased in a way that's confirmatory of the user's existing beliefs ''obama born in kenya'', and phasing can also take a side via terminology used by different sides ''gun control'' vs. ''gun rights''~\cite{koutra2015events}. We will also consider \textit{query bias} and fulfill our experiments accordingly.

\section{Sentiment-wise Comparison of Two Search Engines}

\subsection{Document-level Analysis} In document-level sentiment analysis, we obtain a sentiment value for each document from TextBlob without any pre-processing step on documents since TextBlob handles the punctuation itself. This part of the analysis is more coarse-grained compared to  \textit{sentence-level} analysis.

\subsection{Sentence-level Analysis} Differently from the
document-level, in sentence-level sentiment analysis we view each document as a pile of sentences. We split a given document into sentences and obtain polarity value for each of its sentences from TextBlob. Then, in order to obtain a sentiment score for the given document, we compute an average sentiment score by summing the polarity values of all the sentences and dividing it by the number of sentences in the document, which provides us a more fine-grained analysis of the given documents. That is to say, from TextBlob's point of view, in sentence-level analysis each sentence is seen as a document and processed accordingly. For clarification, we have again only one polarity value for each document to compare the retrieved document sets of two search engines, although we compute these document scores in two distinct ways as \textit{document} and \textit{sentence} level.

\subsection{Sentiment Transformation Procedure}
Apart from the sentiment analysis levels to make a consistent comparison among all controversial topics, one needs to think about the connection between the semantic orientation (i.e. conveying sentiment as positive, negative, or neutral) and the perspective of a document in the scope of its controversial topic. For instance, if a document has a positive sentiment score whose controversial topic is \textit{abortion}, the document's perspective tends to be more \textit{liberal}, whereas if a document shows a positive attitude towards \textit{brexit}, its perspective would be more close to \textit{conservative}. Therefore, to interpret the sentiment score of a document on conservative/liberal scale and use it for comparison, it is necessary to consider sentiment of the given document along with the perspective information associated with its controversial topic. 

Based on the example of interpreting semantic orientation in the context of its controversial topic, after obtaining document polarity values at two different levels via TextBlob, we applied transformation on the document polarity values of some controversial topics. Transformation procedure is employed in order to put controversial topics on the same scale and make a consistent comparison. To illustrate that if there exist two documents and one of them supports \textit{liberal-perspective} towards \textit{abortion} and the other one advocates \textit{brexit}, they will possess different sentiment orientations (i.e. being \textit{liberal} means a positive attitude in the context of \textit{abortion}, whereas it conveys a negative sentiment towards \textit{brexit}). Thus, we need to apply transformation to one of these documents for consistency. 

In this specific example, we transform the \textit{brexit} document and multiply its polarity value by -1. This is because in our evaluation scheme, we prefer to favor liberal documents more, i.e. a more relevant document means a more liberal document for its controversial query, thus we transform the polarity of a negative document if it supports a \textit{liberal-perspective} towards its controversial topic. 
Since transformation is applied after document polarities are computed, there is no difference in \textit{document-level} or \textit{sentence-level} transformation procedure. Note that the transformed topics are specified in Table \ref{table:topdistFirst}, and Table \ref{table:topicsSecond} for two datasets.

\subsection{Comparison Metrics}
We use three different sentiment-based metrics to compare returning documents of search engines. As the first metric, we directly use the average sentiment polarities obtained from TextBlob. Secondly, we propose a NDCG-Senti metric which blends with sentiment and ranking information. Lastly, as the third metric, we use average precision metric which favors the ranked lists containing more frequent relevant documents (\textit{liberal} documents in our scheme) for a given controversial topic.

\subsubsection{Average Sentiment Polarity}
Our first metric compares two search engines by purely using sentiment scores returned from TextBlob and the range for polarity values is [-1, 1]. We obtain sentiment scores for documents, and sentences directly from TextBlob, then compute an average polarity value as mentioned in the previous section for each document by exploiting \textit{document} and \textit{sentence} sentiment values separately. That means that we only care about document sentiment scores for comparison purposes, yet by computing these document scores in two different levels of sentiment analysis, then transforming some of these polarity values if needed. 


\subsubsection{NDCG-Senti}
In addition to the \textit{average sentiment polarity} metric, we also need a more informative metric that will be convenient to compare particularly the retrieved document sets of two search engines. For this reason, we introduce a new metric, \textit{NDCG-Senti} that blends ranking and sentiment information, which is a variant of commonly used NDCG metric. We utilize the traditional formula of NDCG to constitute our ranking-sentiment metric \textit{NDCG-Senti} replacing relevance scores \textit{$rel_i$} with sentiment scores \textit{$pol_i$} in the formula.

With a slight modification in the traditional formula of NDCG, we come up with a modified NDCG scoring function which also takes document rank into consideration in comparing the retrieved document sets of search engines.
However, there is still an issue in the computation of transformed NDCG function that needs to be considered that relevance grades in the traditional NDCG metric are always positive, whereas the real sentiment values as well as the transformed versions of these scores become negative since the polarity values obtained from TextBlob lie between -1 and 1. Therefore, the polarity values of documents are firstly transformed and then normalized to be used for the computation of this secondly proposed metric. In short, we compute \textit{NDCG-Senti} scores with transformed and subsequently normalized sentiment scores of TextBlob. The normalization procedure is described below.




\subsubsection{Average Precision}
Our third comparison metric is \textit{average precision} which computes the frequency of relevant documents in a document list by taking into account of ranking information. In our presented scheme, precision is computed by using the frequency of \textit{liberal} documents (i.e. they are seen as relevant documents in our framework) since we favor \textit{liberal} documents and encourage them to appear in lower ranks. For computation, we used the definition of the expected value of average precision formula proposed in \cite{aslam2005maximum}.

\subsubsection{Fairness}
In our evaluation scheme, we compare search engines mainly in two ways. First of all, we investigate if they both return \textit{fair} (no bias) results and subsequently, if they return sentiment-wise similar set of documents for the same controversial query. However, to evaluate the fairness of the search results retrieved from a search engine is challenging; we need to make a definition of being \textit{fair} mathematically. Or, we should have a baseline \textit{fairness metric} to make a comparison. To introduce a \textit{fairness metric} we need to define the \textit{fair} version of our three proposed sentiment-based comparison metrics, then we can make a comparison between the \textit{fair} metric score and the score computed from a given set of documents. In this way, we can measure if the returning results of a search engine significantly deviate from a \textit{fair} set of documents, using the corresponding comparison metric (average sentiment polarity, NDCG-Senti, or average precision).

\begin{itemize}
    \item \underline{\textit{Fair} Average Sentiment Polarity}: For a \textit{fair} set of documents, this metric should get a value of 0, since the sentiment scores obtained from TextBlob lie in the range of [-1,1].
    \item \underline{\textit{Fair} NDCG-Senti}: To compute the \textit{fair} version of NDCG-Senti, we randomly generate 50 lists of document scores and compute an NDCG score for each of these lists, then
    obtain an average score over these NDCG scores. Note that each generated list contains 10 documents and we randomly choose 0 or 1 for each document score in the list. This is because, NDCG-Senti scores are computed from the normalized sentiment scores and the values lie between 0 and 1. After this procedure, we create a \textit{fair} NDCG-Senti score and we can see if a given document set is \textit{fair} or not using this metric.
    \item \underline{\textit{Fair} Average Precision}:
    To generate the \textit{fair} version of average precision metric, we use the same 50 lists of documents above and compute an average precision score for each of these lists, then calculate an average score over these document sets.
\end{itemize}

\subsubsection{Normalization Procedure}
In order to get rid of negative sentiment scores, we utilize min-max normalization technique to map the polarities onto the range of [0, 1]. For each query, we compute a minimum and a maximum value from the computed polarity values of the document set of each query. Then, we normalize the polarity values using the minimum, maximum values and obtain non-negative scores for the given set of documents. These values are used to compute \textit{NDCG-Senti} and \textit{average precision} for a given set of documents.

\section{Experimental Results}

\subsection{Dataset Crawling}
We collected two datasets using news API of the corresponding search engines: Bing and Google. For this purpose, we implemented scripts on python to automate the crawling process with the API keys of the search engines. We crawled document contents returned by a search engine in response to the controversial topics selected from the website of \url{https://www.procon.org/} in which contains up-to-date controversial issues. We deliberately selected queries as \textit{controversial} in our evaluation scheme to measure bias, if exists and observe the differences between the retrieval results of two popular search engines clearly. We note that the data collection process was fulfilled in a controlled environment such that the same queries were sent to the search engines almost at the same moment since the documents to be retrieved may vary by time.

\subsubsection{Preliminary Dataset}
For the preliminary dataset, we chose 15 "highly" controversial topic from \url{https://www.procon.org/} on the 25th of October, 2017 to see if our proposed method is useful to measure bias and compare the returning document lists of search engines. In this way, we have a rather small dataset and expect more distinguished results of two search engines since we selected the most controversial topics in the website. Since we have a small number of topics, to retrieve more documents related to each topic, we extended the query set by using Google Trends. Then, we filtered the expanded query set, if i) an expanded query does not contain its controversial topic words, or ii) it comprises any sentiment-baring words. In filtering out the first type of extended queries enables us to retrieve more relevant documents about the given controversial topic without losing context information. On the other hand, the second type of filtering helps us on eliminating queries from our query set that possibly introduce bias on search results (i.e. some keywords that may convey positive/negative opinion about its controversial topic), thus being able to focus on the contents of the retrieved documents for a more \textit{fair} comparison analysis. In this way, we aim to send extensive but \textit{objective} queries to a search engine to obtain related documents about all the aspects of a controversial topic.

For each query sent to the search engine, we crawled the first 10 documents returned from that search engine. The dataset is composed of 2030 documents in total for one search engine. The topic distribution of the crawled dataset is shown in Table \ref{table:topdist}. In the table, one can see that there are 160 documents crawled only for the controversial topic of \textit{abortion}. This means that we have 16 expanded queries for \textit{abortion} since we use only the first 10 documents retrieved from the corresponding search engine. Moreover, topics whose sentiment scores will be transformed before computing the metrics are bold-faced in the table. Topics to be transformed were determined by the majority voting of human annotators and topics whose sides are clearly certain on conservative-liberal scheme were selected for transformation. 

After having expanded the query set, we applied filtering process on the extended query sets of all controversial topics in the dataset, and almost all query set sizes decreased. One may argue that some query sets are really small in number as shown in Table \ref{table:topdistFirst}. However, note that small query set sizes were not the outcome of the filtering process since the process did not exclude high number of queries from the query sets. Small query set sizes may be caused by the fact that the initial version of these extensive query sets at the beginning were also very small comparing to the rest. On the other hand, one reason for having small sized initial query sets could be less number of various aspects/subtopics in the corresponding controversial topic to get attention and being discussed on the Internet.

\begin{table}[!ht]
\begin{center}
\begin{tabular}{ |c|c|c| } 
 \hline
 \textbf{Topic} & \textbf{\# of documents} \\ 
 \hline
 Abortion & 160 \\ 
 \hline
 \textbf{Animal Testing} & 60  \\ 
 \hline
 Assisted Suicide  & 70 \\ 
 \hline
 \textbf{Brexit} & 150 \\ 
 \hline
 Climate Change & 300 \\ 
 \hline
 Gay Marriage & 180 \\ 
 \hline
 Gun Control & 200 \\ 
 \hline
 Medical Marijuana  & 130 \\ 
 \hline
 \textbf{Minimum Wage}  & 220 \\ 
 \hline
 Obamacare  & 60 \\ 
 \hline
 Prostitution & 40 \\ 
 \hline
 Syrian Refugees & 80 \\ 
 \hline
 Transgender Military  & 60 \\ 
 \hline
 \textbf{Travel Ban} & 90 \\ 
 \hline
 \textbf{Trump} & 230 \\ 
 \hline
\end{tabular}
\end{center}
\caption{\textit{Preliminary} Dataset - 15 Controversial Topic Distributions (transformation was applied on bold-faced topics)}
\label{table:topdistFirst}
\end{table}

\subsubsection{Stable Dataset}
In the second dataset, for a more comprehensive analysis, we got all 62 controversial topics in the website of \url{https://www.procon.org/} on the 31th of July, 2018. Our \textit{stable} dataset differs from the \textit{preliminary} dataset essentially in the following elements:

\begin{itemize}
    \item We did not select the controversial topic set since we need a bigger dataset and also not to affect the results by injecting bias on the input.  Because the controversial issues given in the website are mostly about US, we behaved as if our location was somewhere in US while crawling. Differently from the \textit{preliminary} dataset, we crawled ooth the first 10 and 100 returned documents for each controversial query. \\
    
    \item Also, we did not expand the query set in this dataset because of the issues encountered in establishing the first one and it seems to prevent us from creating a controlled environment by possibly introducing bias on the input. Thus, we used 62 controversial issues presented in the website exactly as they are (i.e. including uppercased characters, and without removing punctuations). Note that since we did not employ query expansion in the second dataset, controversial topics are our queries at the same time, so we can use a controversial topic and a controversial query interchangeably for the second dataset. The controversial queries for the \textit{stable} dataset are shown and the topics whose sentiment scores will be transformed are bold-faced in the Table \ref{table:topicsSecond}. These topics were determined by the majority voting of human annotators and the ones whose sides are clearly certain on conservative-liberal scheme were selected for transformation.  \\

    \item Since there is no query expansion phase, we have one list of documents for each controversial query (topic). Hence, we have a single NDCG-senti, and an average precision score computed for each topic so statistical significance test can be applied on these computed scores directly. For average sentiment polarity, on the other hand, there is one sentiment score for each document in the given set, thus we need to compute an average of these scores to obtain a single score for each topic to be compared. However, in the case of the first dataset, there exists one NDCG-Senti, and average precision score for each query, since each controversial topic consists of many queries, we need to compute an average of these scores to generate one single score for each controversial topic. Additionally, for average sentiment polarity, we initially computed an average over all the document polarities in the given document set for one expanded query, subsequently we needed to compute the mean of these averaged sentiment polarities to obtain one mean average sentiment polarity for comparison.  \\
    
    \item Lastly, we also made a query analysis and obtained the sentiment polarities for the controversial queries from TextBlob. Polar queries are displayed in Table \ref{table:topicsSecondPolar} with their polarity information. We repeated the same analyses with and without these non-neutral queries to remove a possible source of bias rooted from queries themselves. 
\end{itemize}

With these four factors, we aimed to crawl a more stable dataset in a more controlled environment without introducing any bias on the input and to analyse the contents of the documents returned from search engines in terms of sentiment.

\begin{table}[!ht]
\small
\begin{center}
\resizebox{\columnwidth}{!}{%
\begin{tabular}{|l|}
 \hline
 \textbf{Topic Names (three topics on each line)} \\ 
 \hline
 2016 Presidential Election, Abortion, ACLU - Good for America? \\
 \hline
 Alternative Energy vs. Fossil Fuels, \textbf{Animal Testing}, \textbf{Banned Books} \\
\hline
Bill Clinton, Born Gay? Origins of Sexual Orientation, Bottled Water Ban \\
\hline
Cell Phone Radiation?, \textbf{Churches and Taxes}, Climate Change \\
\hline
College Education Worth It?, \textbf{Concealed Handguns}, \textbf{Corporal Punishment} \\
\hline
\textbf{Corporate Tax Rate \& Jobs}, \textbf{Cuba Embargo}, Daylight Saving Time \\
\hline
Death Penalty, Drinking Age - Lower It?, Drone Strikes Overseas \\
\hline
Drug Use in Sports, \textbf{Electoral College}, Euthanasia \& Assisted Suicide \\
\hline
Felon Voting, Fighting in Hockey, Gay Marriage \\
\hline
\textbf{Gold Standard}, Golf - Is It a Sport?, Gun Control \\
\hline
Illegal Immigration, Israeli-Palestinian Conflict, Medical Marijuana \\
\hline
Milk - Is It Healthy?, \textbf{Minimum Wage}, National Anthem Protests \\
\hline
Net Neutrality, Obamacare, Obesity a Disease? \\
\hline
Olympics, Penny - Keep It?, Police Body Cameras \\
\hline
Prescription Drug Ads, Prostitution - Legalize It?, Recreational Marijuana \\
\hline
Right to Health Care, \textbf{Ronald Reagan}, Sanctuary Cities \\
\hline
\textbf{School Uniforms}, \textbf{School Vouchers}, Social Media? \\
\hline
\textbf{Social Security Privatization}, Standardized Tests, \textbf{Student Loan Debt} \\
\hline
Tablets vs. Textbooks, Teacher Tenure, Under God in the Pledge \\
\hline
Universal Basic Income, Vaccines for Kids, Vegetarianism \\
\hline
Video Games and Violence, Voting Machines \\
\hline
\end{tabular}
}
\end{center}
\caption{\textit{Stable} Dataset - 62 Controversial Topics (transformation was applied on bold-faced topics)}
\label{table:topicsSecond}
\end{table}

\begin{table}[!ht]
\small
\begin{center}
\begin{tabular}{ |c|c|c| } 
 \hline
 \textbf{Topic} & \textbf{Polarity}\\ 
 \hline
ACLU - Good for America?	&	0.70\\	\hline
Born Gay? Origins of Sexual Orientation	&	0.46	\\	\hline
College Education Worth It?	&	0.30	\\	\hline
Gay Marriage	&	0.42\\	\hline
Illegal Immigration	&	-0.50\\	\hline
Milk - Is It Healthy?	&	0.50	\\	\hline
Right to Health Care	&	0.29	\\	\hline
Social Media?	&	0.03	\\	\hline
Social Security Privatization 	&	0.03\\	\hline

\end{tabular}
\end{center}
\caption{Polar Controversial Queries In the \textit{Stable} Dataset}
\label{table:topicsSecondPolar}
\end{table}

\subsection{Evaluation Framework}
In evaluating search engines, our main intent is to compare these two search engines on conservative/liberal perspective towards a given controversial topic. For comparison, we utilized TextBlob sentiment scores at two different levels as \textit{document} and \textit{sentence}. For evaluation, we used three sentiment-based metrics and computed these at two different levels, \textit{document} and \textit{sentence} on the returning documents of Bing and Google. In using these metrics, we aimed to answer the two following research questions we raised in the beginning of the paper. 

\textbf{RQ1: Do search engines return \textit{non-biased} result sets in response to queries related to controversial issues?}                 

To find an answer to this question, we need to compare the retrieved results of a search engine with a \textit{fair} list of documents. For this purpose, we computed the scores of three proposed comparison metric scores for all the retrieved results of a search engine and the scores of \textit{fair} version of the same metrics. Then we applied a two-tail statistical significance test on these scores for each metric (i.e. average sentiment polarity, NDCG-Senti, and average precision) to see if there is a significant difference between the retrieved results of a search engine and a \textit{fair} list of documents. In this way, we can decide if a search engine returns \textit{fair} results in terms of the corresponding metric.

\textbf{RQ2: On conservative-liberal scale, are the viewpoints of search engines \textit{significantly different} from each other towards controversial topics?}

Subsequently, to investigate and answer to the second question, we need to compare the overall perspectives of Bing and Google on the same controversial topics. For this reason, we computed three comparison metric scores on both Bing and Google's returning document lists for all the controversial queries in the query set. Then, we applied a two-tail statistical significance test on these values separately for each comparison metric and see if there is a significant difference between Bing and Google in terms of the corresponding metric.

\subsubsection{Comparison Analysis}
To answer the research questions
above, we used two datasets. The first dataset was used only for \textit{preliminary} results as its name implies. More comprehensive analysis was done in the \textit{stable} dataset which was prepared in a more controlled environment, that contains the first 10 as well as 100 retrieved documents and also incorporates query analysis to avoid input bias.

\paragraph{\textit{Stable} Dataset Results}: The raised research questions will be answered for the second dataset in a more comprehensive manner.

\underline{All controversial queries}: As you can see from the Table \ref{table:sigPol_10Second}, \ref{table:sigNDCG_10Second}, and \ref{table:sigAP_10Second}, all three comparison metrics gave consistent results. The computed scores demonstrate that Bing and Google are similar but they do not seem \textit{fair}. Moreover, average sentiment polarity scores computed from the first 100 documents also support the same claim. Therefore, this may mean that not ranking algorithms of search engines, but rather the corpus for the controversial issues are \textit{biased}.

\begin{table}[!h]
\small
\begin{center}
\begin{tabular}{ |c|c|c|c| } 
 \hline
 \textbf{Comparison} & \textbf{Analysis Level} & \textbf{p-value}  &  \textbf{\textit{Significant?}}\\ 
 \hline
  Bing-FAIR & Doc & 0.00192 &	\textbf{YES}\\ 
 \hline
  Google-FAIR &  Doc & 0.00095 &	\textbf{YES}\\ 
 \hline
  Bing-Google & Doc & 0.41807 &	\textbf{NO}\\ 
 \hline
  Bing-FAIR &  Sent & 0.00134 &	\textbf{YES}\\ 
 \hline
  Google-FAIR & Sent & 0.00097 &	\textbf{YES}\\ 
  \hline
  Bing-Google & Sent & 0.74972 &	\textbf{NO}\\ 
 \hline
\end{tabular}
\end{center}
\caption{\textit{Stable} Dataset - two-tail t-test ($\alpha$:0.01) on average sentiment polarity@10}
\label{table:sigPol_10Second}
\end{table}

\begin{table}[!h]
\small
\begin{center}
\begin{tabular}{ |c|c|c|c| } 
 \hline
 \textbf{Comparison} & \textbf{Analysis Level} & \textbf{p-value}  &  \textbf{\textit{Significant?}}\\ 
 \hline
  Bing-FAIR & Doc & 1.4E-09&	\textbf{YES}\\ 
 \hline
  Google-FAIR &  Doc & 2.0E-07&	\textbf{YES}\\ 
 \hline
  Bing-Google & Doc & 0.9259 &	\textbf{NO}\\ 
 \hline
  Bing-FAIR &  Sent & 4.8E-09 &	\textbf{YES}\\
 \hline
  Google-FAIR & Sent & 1.7E-08 &	\textbf{YES}\\ 
  \hline
  Bing-Google & Sent & 0.6696 &	\textbf{NO}\\ 
 \hline
\end{tabular}
\end{center}
\caption{\textit{Stable} Dataset - two-tail t-test ($\alpha$:0.01) on NDCG@10}
\label{table:sigNDCG_10Second} 
\end{table}

\begin{table}[!h]
\small
\begin{center}
\begin{tabular}{ |c|c|c|c| } 
 \hline
 \textbf{Comparison} & \textbf{Analysis Level} & \textbf{p-value}  &  \textbf{\textit{Significant?}}\\ 
 \hline
  Bing-FAIR & Doc & 0.00107 & \textbf{YES}\\
 \hline
  Google-FAIR &  Doc & 0.00480 &\textbf{YES}\\
 \hline
  Bing-Google & Doc & 0.92069 &	\textbf{NO}\\ 
 \hline
  Bing-FAIR &  Sent & 0.00196 &	\textbf{YES}\\
 \hline
  Google-FAIR & Sent & 0.00353 &\textbf{YES}\\
  \hline
  Bing-Google & Sent & 0.89756 &	\textbf{NO}\\ 
 \hline
\end{tabular}
\end{center}
\caption{\textit{Stable} Dataset - two-tail t-test ($\alpha$:0.01) on AP@10}
\label{table:sigAP_10Second} 
\end{table}

\begin{table}[!h]
\small
\begin{center}
\begin{tabular}{ |c|c|c|c| } 
 \hline
 \textbf{Comparison} & \textbf{Analysis Level} & \textbf{p-value}  &  \textbf{\textit{Significant?}}\\ 
 \hline
  Bing-FAIR & Doc & 0.00051&	\textbf{YES}\\ 
 \hline
  Google-FAIR &  Doc & 0.00221&	\textbf{YES}\\ 
 \hline
  Bing-Google & Doc & 0.75505 &	\textbf{NO}\\ 
 \hline
  Bing-FAIR &  Sent & 0.00086&	\textbf{YES}\\ 
 \hline
  Google-FAIR & Sent & 0.00248&	\textbf{YES}\\ 
  \hline
  Bing-Google & Sent & 0.77645&	\textbf{NO}\\ 
 \hline
\end{tabular}
\end{center}
\caption{\textit{Stable} Dataset - two-tail t-test ($\alpha$:0.01) on average sentiment polarity@100}
\label{table:sigPol_100}
\end{table}

\begin{figure}[!h]
\centering
\includegraphics[width=\columnwidth]{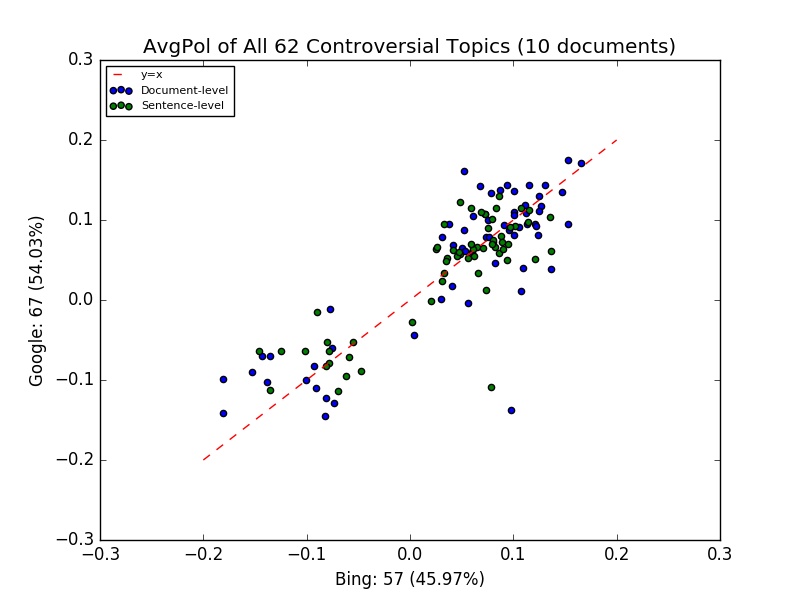}
\caption{\textit{Stable} Dataset - controversial topic distribution on average sentiment polarity@10 scores}
\label{fig:avgPol_10}
\end{figure}

\begin{figure}[!h]
\centering
\includegraphics[width=\columnwidth]{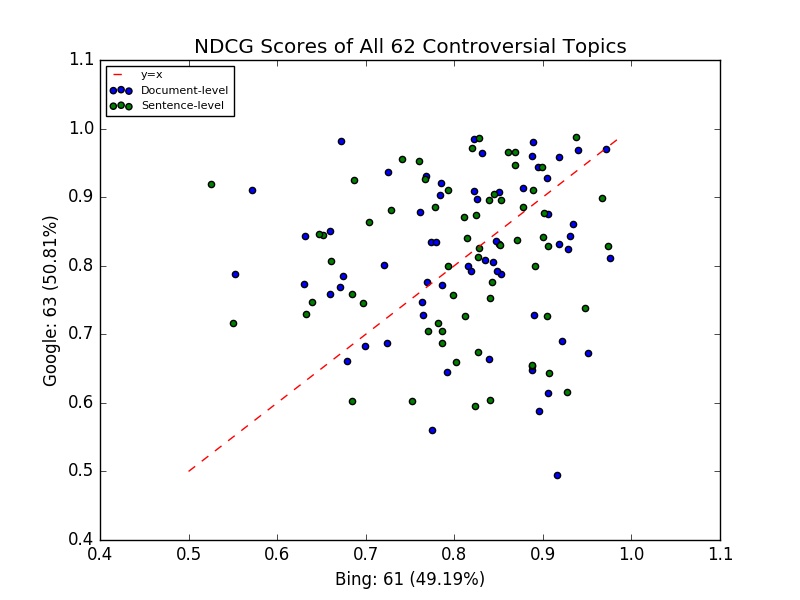}
\caption{\textit{Stable} Dataset - controversial topic distribution on NCDG@10 scores}
\label{fig:NDCG_10}
\end{figure}

\begin{figure}[!h]
\centering
\includegraphics[width=\columnwidth]{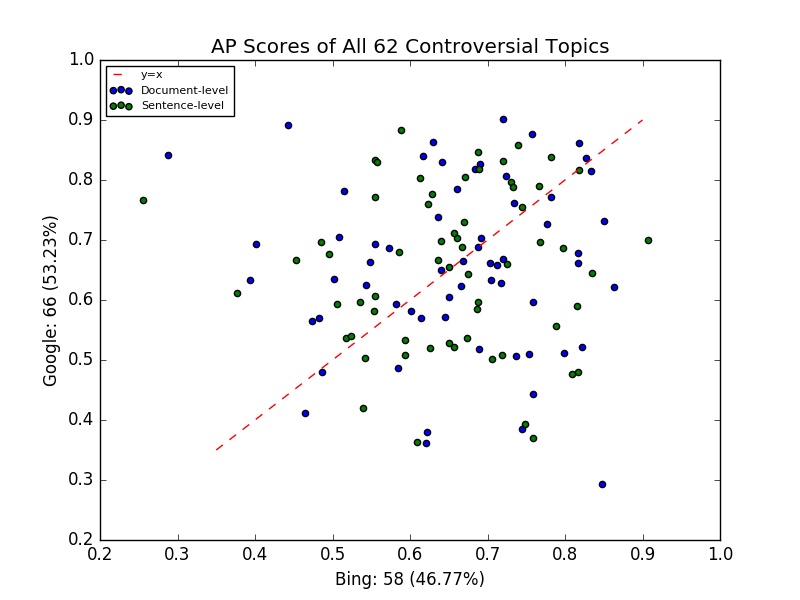}
\caption{\textit{Stable} Dataset - controversial topic distribution on AP@10 scores}
\label{fig:AP_10}
\end{figure}

\begin{figure}[!h]
\centering
\includegraphics[width=\columnwidth]{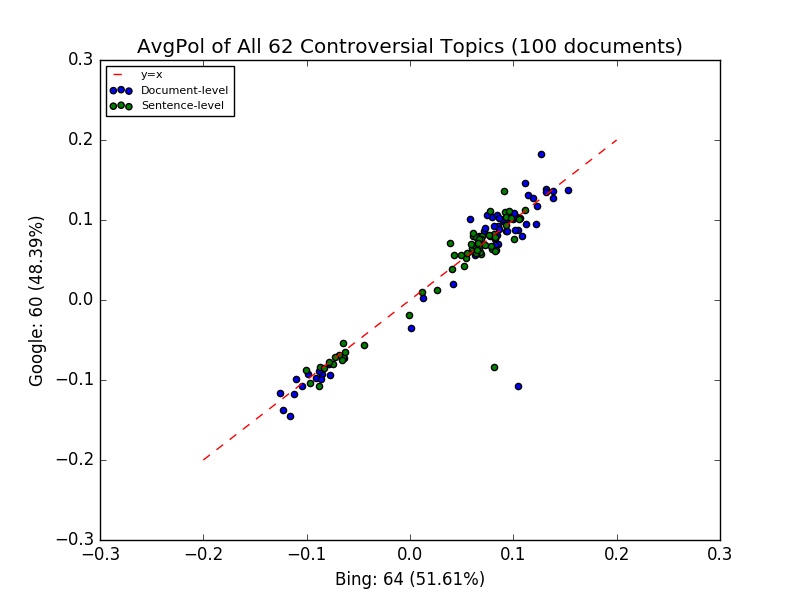}
\caption{\textit{Stable} Dataset - controversial topic distribution on average sentiment polarity@100 scores}
\label{fig:avgPol_100}
\end{figure}

Moreover, we plotted all the controversial topics to visualize the distribution of average sentiment polarity, NDCG-Senti, and average precision scores for the analysis of the first 10 documents, and average sentiment polarity values for the first 100 documents in Figure \ref{fig:avgPol_10}, \ref{fig:NDCG_10}, \ref{fig:AP_10} and \ref{fig:avgPol_100}. In these plots, we denoted each controversial query as a point; blue for a document-level and green for a sentence-level score and put a y=x line (i.e. scores of Bing and Google are equal on the line) for comparison. If a point is on the above of the line, then its Google score is more positive than Bing and vice-versa. You can also find the number of points with their percentages on the x and y axis label of the corresponding plot.

\underline{Only non-polar controversial queries}: In this part, we removed
the polar queries that are displayed in Table \ref{table:topicsSecondPolar}. In this way, we aimed to eliminate input bias that may be caused from these polar queries and obtain more accurate results. One can defend that some of the polar queries shown in the Table 
\ref{table:topicsSecondPolar} may not look like polarized queries, however we did not want to pick the polar ones manually and affect the results. Therefore, in our procedure, we accepted a given query as polar if the TextBlob assigns a sentiment polarity that is different than 0 to it. Since we also obtained the sentiment scores from the TextBlob that we used in our framework, the whole comparison procedure is consistent in itself, although some queries do not seem strongly polarized.

Experiments fulfilled with non-polar queries only, the same result can be
drawn from the average sentiment scores for the first 10 and 100 documents that most probably the corpus is biased (i.e. there is input bias) , rather than the ranking algorithms of search engines as seen from Table \ref{table:sigPol_10NonPolar} and \ref{table:sigPol_100NonPolar}. Yet differently from the previous dataset, in the average sentiment scores for the first 10 and 100 documents, Bing and Google seem to be fair with the confidence of 99\% if $\alpha$ is set to 0.05, then these scores also support the same claim as the scores of the previous dataset did. Thus, one can advocate that if we eliminate the polar queries from the query set, in terms of average sentiment polarity, both search engines seem to return more \textit{fair} results which encourages the fact that we could have eliminated the injected input bias when we eliminated the polar queries.

\begin{table}[!h]
\small
\begin{center}
\begin{tabular}{ |c|c|c|c| } 
 \hline
 \textbf{Comparison} & \textbf{Analysis Level} & \textbf{p-value}  &  \textbf{\textit{Significant?}}\\ 
 \hline
  \textbf{Bing-FAIR} & Doc & \textbf{0.01704*} &	\textbf{NO}\\ 
 \hline
  Google-FAIR &  Doc & 0.00631&	\textbf{YES}\\
 \hline
  Bing-Google & Doc & 0.27039 &	\textbf{NO}\\ 
 \hline
  \textbf{Bing-FAIR} &  Sent & \textbf{0.01178*} &	\textbf{NO}\\ 
 \hline
  Google-FAIR & Sent & 0.00721 &	\textbf{YES}\\ 
  \hline
  Bing-Google & Sent & 0.46517 &	\textbf{NO}\\ 
 \hline
\end{tabular}
\end{center}
\caption{\textit{Non-Polar Stable} Dataset - two-tail t-test ($\alpha$:0.01) on average sentiment polarity@10}
\label{table:sigPol_10NonPolar}
\end{table}

\begin{table}[!h]
\small
\begin{center}
\begin{tabular}{ |c|c|c|c| } 
 \hline
 \textbf{Comparison} & \textbf{Analysis Level} & \textbf{p-value}  &  \textbf{\textit{Significant?}}\\ 
 \hline
  Bing-FAIR & Doc & 8.7E-08 &	\textbf{YES}\\ 
 \hline
  Google-FAIR &  Doc & 2.9E-07 &	\textbf{YES}\\ 
 \hline
  Bing-Google & Doc & 0.78749 &	\textbf{NO}\\ 
 \hline
  Bing-FAIR &  Sent & 1.7E-06 &	\textbf{YES}\\ 
 \hline
  Google-FAIR & Sent & 7.6E-10 &	\textbf{YES}\\ 
  \hline
  Bing-Google & Sent & 0.44474 &	\textbf{NO}\\ 
 \hline
\end{tabular}
\end{center}
\caption{\textit{Non-Polar Stable} Dataset - two-tail t-test ($\alpha$:0.01) on NDCG@10}
\label{table:sigNDCG_10NonPolar} 
\end{table}

\begin{table}[!h]
\small
\begin{center}
\begin{tabular}{ |c|c|c|c| } 
 \hline
 \textbf{Comparison} & \textbf{Analysis Level} & \textbf{p-value}  &  \textbf{\textit{Significant?}}\\ 
 \hline
  Bing-FAIR & Doc & 0.00586 &	\textbf{YES}\\ 
 \hline
  Google-FAIR &  Doc & 0.00642 &	\textbf{YES}\\ 
 \hline
  Bing-Google & Doc & 0.85482&	\textbf{NO}\\ 
 \hline
  Bing-FAIR &  Sent & 0.01378 &	\textbf{YES}\\ 
 \hline
  Google-FAIR & Sent & 0.00241 &	\textbf{YES}\\ 
  \hline
  Bing-Google & Sent & 0.69829&	\textbf{NO}\\ 
 \hline
\end{tabular}
\end{center}
\caption{\textit{Non-Polar Stable} Dataset - two-tail t-test ($\alpha$:0.01) on AP@10}
\label{table:sigAP_10NonPolar} 
\end{table}

\begin{table}[!h]
\small
\begin{center}
\begin{tabular}{ |c|c|c|c| } 
 \hline
 \textbf{Comparison} & \textbf{Analysis Level} & \textbf{p-value}  &  \textbf{\textit{Significant?}}\\
 \hline
  Bing-FAIR & Doc & 0.00684&	\textbf{YES}\\ 
 \hline
  \textbf{Google-FAIR} &  Doc & \textbf{0.02263*}&	\textbf{NO}\\ 
 \hline
  Bing-Google & Doc & 0.77357 &	\textbf{NO}\\ 
 \hline
  Bing-FAIR &  Sent & 0.00986&	\textbf{YES}\\ 
 \hline
  \textbf{Google-FAIR} & Sent & \textbf{0.02540*}&	\textbf{NO}\\ 
  \hline
  Bing-Google & Sent & 0.84905&	\textbf{NO}\\ 
 \hline
\end{tabular}
\end{center}
\caption{\textit{Non-Polar Stable} Dataset - two-tail t-test ($\alpha$:0.01) on average sentiment polarity@100}
\label{table:sigPol_100NonPolar}
\end{table}

\section{Conclusion and Future Work}
To sum up, our experiments support that both commercial search engines seem to return \textit{biased} results for controversial queries. Since we also measured bias in the corpus containing the first 100 documents, the bias may also be caused by the documents in the corpus for the given controversial query. Moreover, from an overall perspective, Bing and Google seem to have similar points of view on the same controversial topic.

As a future work, we plan to apply aspect-level sentiment analysis on the retrieved documents for a more detailed framework. In this way, we can analyse the search results in a more fine-grained manner and it is necessary, especially for the controversial queries including two sides in its textual content such as "Alternative Energy vs. Fossil Fuels" and "Israeli-Palestinian Conflict", etc. for a proper comparison. Furthermore, documents can be retrieved from the search engines of different countries and a location-based bias analysis can be incorporated to our existing framework.



\newpage

\bibliographystyle{ACM-Reference-Format}
\bibliography{main}

\end{document}